\documentclass[]{aa}

\usepackage{natbib}
\usepackage{breqn}
\usepackage{xcolor}
\usepackage{graphicx}
\usepackage{txfonts}
\usepackage{multirow}

\usepackage{hyperref} 

\hypersetup{
    colorlinks = true,
    linkbordercolor = {cyan},
    linkcolor={blue},
    citecolor={blue},}

\begin{document}

\def\figureautorefname{Fig.}
\def\equationautorefname~#1\null{Eq. (#1)\null}

\def\sectionautorefname{Sect.}
\def\subsectionautorefname{Sect.}
\def\subsubsectionautorefname{Sect.}

\title{Prescription for the rotational and orbital evolution of main-sequence binaries undergoing resonance locking}
\titlerunning{MAD Tides}

\author{L. Fellay\inst{1} \and M.-A. Dupret\inst{1} }
\institute{STAR Institute, University of Liège, 19C Allée du 6 Août, B$-$4000 Liège, Belgium}
\date{June,  2026}

\abstract{
In intermediate-mass binary stars, dynamical tides may play an important role in the orbital and rotational evolution of the system. One possible channel for long-term angular-momentum exchange is resonance locking, in which the evolution of the orbital and rotational parameters remains coupled to the evolution of the frequency of a resonantly excited oscillation mode.
}
{
We aim to develop a simple prescription for describing the orbital and rotational evolution of binary systems undergoing resonance locking, and to assess the range of validity of this prescription.
}
{
Using the asymptotic formalism for high-order gravity modes, we relate the temporal evolution of mode frequencies to the evolution of the orbital and rotational properties of the binary system. We then compare the resulting prescription with full integrations of the dynamical-tide evolution, including the contribution of multiple forcing frequencies.
}
{
Depending on the type of resonance locking considered, we identify the regions of parameter space in which locking is expected to be possible in terms of orbital separation and rotation rate. We find that resonance locking is favoured in systems with significant spin--orbit asynchronism. When a locking is maintained, the proposed prescription reproduces the main features of the orbital and rotational evolution obtained from full dynamical-tide integrations, without requiring the explicit computation of the complete forcing-frequency spectrum.
}
{
If resonance locking occurs, the evolution of the binary system can be described by a relatively simple prescription. Comparison with the observed population of binaries containing \(\gamma\)-Doradus stars suggests that a significant fraction of the sample lies in regions compatible with the expected location of locked systems. However, improved characterisation of these systems is required to distinguish resonance locking from alternative explanations, such as youth, initially asynchronous rotation, or weak tidal interactions.
}
\maketitle
\noindent

\section{Introduction}
In intermediate-mass stars, the orbital and rotational evolution of binary systems can be strongly affected by dynamical tides, which arise from the resonant excitation of stellar oscillation modes by the tidal potential \citep{Zahn1970,Zahn1975,Zahn1977,Smeyers1998,Willems2003}. In the low-frequency regime usually considered by \citet{Zahn1975,Zahn1977}, where resonances with individual low-order modes are not resolved, the dynamical tide acts as a dissipative process. Its effect can then be described through the corresponding tidal-dissipation formalism, leading in general to the exchange of angular momentum between the orbit and stellar rotation and driving the system towards synchronisation and circularisation \citep{Hut1981,Goldreich1989}. Outside this regime, resonances with individual oscillation modes can become important \citep{Witte1999a,Fuller2012,Fuller2017,Fellay2025,gyre_tides2023}. 
\\~\\Resonances with stable modes, which are damped inside the star, dissipate orbital energy and therefore produce the usual tidal behaviour. By contrast, resonances with unstable modes can give rise to the opposite effect: energy supplied by the mode-driving mechanism can be transferred to the orbit or to stellar rotation, potentially increasing the orbital energy and driving the system away from synchronisation. This phenomenon is referred to as inverse tides \citep{Fuller2021,Fellay2026}. Inverse tides can therefore occur only when unstable modes are present. They are expected in stars located within instability strips, where a driving mechanism feeds energy into the oscillations. In this work, we focus on \(\gamma\)-Doradus stars.
\\~\\The \(\gamma\)-Doradus stars are late-A- to F-type main-sequence pulsators, with typical masses of approximately \(1.4\)--\(2.0\,M_{\odot}\) \citep{Li2020}. Their high-order gravity modes are driven by convective-flux blocking near the base of the convective envelope \citep{Guzik2000,Dupret2005,Fellay2026b}. Although their rotation rates span a broad range, most stars with both near-core and surface rotation measurements are consistent with approximately solid-body rotation \citep{Li2020}. Their observed gravity-mode periods are typically of the order of a day \citep{Li2020b}, comparable to the day-scale orbital periods found in close binary systems \citep{Li2020b}.
\\~\\In a recent study, we developed an orbital-evolution code that consistently accounts for the effects of dynamical tides associated with multiple forcing frequencies throughout the evolution of a binary system \citep{Fellay2026}. We applied this code to \(\gamma\)-Doradus binaries and investigated the role of inverse tides. Our results suggest that inverse tides can significantly affect the evolution of these systems when coupled to resonance locking, and that such locking may remain stable over long timescales. Resonance locking occurs when the tidal exchange of angular momentum causes the forcing frequency to evolve at the same rate as the frequency of the resonantly excited mode, thereby maintaining the resonance. Since the oscillation frequencies in \(\gamma\)-Doradus stars increase with time, the torque required to maintain resonance locking has the sign associated with resonances with unstable modes. During resonance locking, the evolution of the orbit and stellar rotation is coupled to that of the mode frequency. The long-term evolution of the system is therefore largely controlled by the frequency evolution of the mode that maintains the locking \citep{Witte1999b,Witte2001,Burkart2012,Burkart2014,Fuller2012,Fuller2017}. Observational candidates for resonance locking have also been identified among heartbeat stars, most notably KIC~8164262 \citep{FullerEtAl2017,Hambleton2018}.
\\~\\Although the amplitude of the dynamical tide generally requires non-adiabatic oscillation calculations when resonances with low-radial-order modes are involved \citep{Willems2003,Fellay2025}, the secular evolution of gravity mode frequencies can be approximated using the asymptotic formalism \citep{Tassoul1980}. The effect of rotation on the oscillation frequencies can be included through the traditional approximation \citep{Lee1997,Townsend2005,Bouabid2013}. These formalisms can therefore be used to construct a simple prescription for the orbital and rotational evolution of binary systems undergoing resonance locking. In this article, we derive such a prescription for the evolution of stellar rotation, orbital eccentricity, and semi-major axis. We then assess its validity by comparing the resulting orbital evolution with full integrations of the dynamical-tide problem, including multiple forcing frequencies.
\\~\\In \autoref{subsect_rot_criterion_resonance_locking}, we present the asymptotic formalism used in this work and derive the prescription for the evolution of the rotational and orbital parameters of binary systems. In \autoref{sect_locking_retrograde}, we then investigate the regimes of rotational and orbital parameters in which resonance locking with retrograde modes is expected to occur, before comparing the prescription with full dynamical-tide integrations. In \autoref{sect_prograde_modes}, we consider resonance locking with prograde modes and discuss the limitations of the prescription in this regime. Finally, in \autoref{sect_discussion}, we examine the range of validity of the prescription in more detail, identify the regions where resonance locking is expected to occur, and compare them with the observed population of close binaries containing \(\gamma\)-Doradus stars. We conclude in \autoref{sect_conclusion}.

\section{Evolution of the rotation rate during resonance locking}\label{subsect_rot_criterion_resonance_locking}

The effects of tidally excited oscillation modes on the orbital and rotational parameters of the system are given by \citep{Smeyers1998,Willems2003,CAFeiN2013,Fellay2025}
\begin{equation}\label{eq_secular_a}
\dfrac{\mathrm{d}a}{\mathrm{d}t} =4\Omega_{\mathrm{orb}}\dfrac{M_2}{M_1}a\sum_{\ell=2}^{\infty} \sum_{m=-\ell}^{\ell} \sum_{k=-\infty}^{\infty} \left( \dfrac{R_1}{a} \right)^{\ell+3} \mathrm{Im}(F_{\ell,m,k}) G^{(2)}_{\ell,m,k}(e),
\end{equation}
\begin{equation}\label{eq_secular_e}
\dfrac{\mathrm{d}e}{\mathrm{d}t} =4\Omega_{\mathrm{orb}}\dfrac{M_2}{M_1}\sum_{\ell=2}^{\infty} \sum_{m=-\ell}^{\ell} \sum_{k=-\infty}^{\infty} \left( \dfrac{R_1}{a} \right)^{\ell+3} \mathrm{Im}(F_{\ell,m,k}) G^{(3)}_{\ell,m,k}(e),
\end{equation}
\begin{align}\label{eq_secular_rot}
\dfrac{1}{I}\dfrac{\mathrm{d}L_{\mathrm{rot}}}{\mathrm{d}t}
&=\dfrac{\mathrm{d}\Omega_{\mathrm{rot}}}{\mathrm{d}t}
+\Omega_{\mathrm{rot}}\dfrac{\mathrm{d}\ln I}{\mathrm{d}t}=4\dfrac{\Omega_{\mathrm{orb}}}{I}
\left(\dfrac{G a}{M_1+M_2}\right)^{1/2} M_2^2\nonumber\\
&\quad\times\sum_{\ell=2}^{\infty} \sum_{m=-\ell}^{\ell} \sum_{k=-\infty}^{\infty}
\left( \dfrac{R_1}{a} \right)^{\ell+3}
\mathrm{Im}(F_{\ell,m,k}) G^{(4)}_{\ell,m,k}(e).
\end{align}
where $M_1$ and $R_1$ are the mass and radius of the primary, respectively, $M_2$ is the mass of the secondary, $a$ is the semi-major axis, and $\Omega_{\mathrm{orb}}$ is the orbital angular frequency. The quantities $L_{\mathrm{rot}}$, $I$, and $\Omega_{\mathrm{rot}}$ are the primary's rotational angular momentum, moment of inertia, and rotational angular frequency, respectively. The quantity $\mathrm{Im}(F_{\ell,m,k})$ provides a dimensionless measure of the stellar response entering the secular evolution equations for each forcing component, characterised by the spherical harmonic degree $\ell$, azimuthal order $m$, and orbital harmonic index $k$ \citep{Smeyers1998,Willems2003,Willems2010,CAFeiN2013}. This is the only factor in these expressions that typically requires evaluation using a non-adiabatic stellar oscillation code. Finally, the coefficients $G^{(2)}_{\ell,m,k}(e)$, $G^{(3)}_{\ell,m,k}(e)$, and $G^{(4)}_{\ell,m,k}(e)$ describe how the orbital geometry weights each stellar response in the evolution equations. Their parenthesised superscripts distinguish the contributions to the evolution of the semi-major axis, eccentricity, and rotational angular momentum, respectively. Explicit expressions for these coefficients can be found in \citet{Smeyers1998,Willems2003,Willems2010,CAFeiN2013,Fellay2026}.
\\~\\The evolution of the stellar rotation rate can be decomposed into a tidal component $\left(\mathrm{d}\Omega_{\mathrm{rot}}/\mathrm{d} t\right)_{\mathrm{tides}}$ and an evolutionary component $\left(\mathrm{d}\Omega_{\mathrm{rot}}/\mathrm{d} t\right)_{\mathrm{evo}}$:
\begin{equation}\label{eq_decomposition_rotation}
\dfrac{\mathrm{d}\Omega_{\mathrm{rot}}}{\mathrm{d} t}=\left(\dfrac{\mathrm{d}\Omega_{\mathrm{rot}}}{\mathrm{d} t}\right)_{\mathrm{tides}}+\left(\dfrac{\mathrm{d}\Omega_{\mathrm{rot}}}{\mathrm{d} t}\right)_{\mathrm{evo}},
\end{equation}
The evolutionary component, which follows from the change in the stellar moment of inertia, is
\begin{equation}
\left(\dfrac{\mathrm{d}\Omega_{\mathrm{rot}}}{\mathrm{d} t}\right)_{\mathrm{evo}}=-\dfrac{\mathrm{d} \ln I}{\mathrm{d} t}\Omega_{\mathrm{rot}},
\end{equation} 
while the tidal component is given by the right-hand side of \autoref{eq_secular_rot}. In general, several forcing components can contribute simultaneously to the orbital evolution \citep{Willems2003,Fuller2017,Fellay2025}. In a binary system, the forcing frequencies in the co-rotating frame can be written as
\begin{equation}\label{eq_forcing_frequency}
\sigma_{m,k}= k\Omega_{\mathrm{orb}} + m\Omega_{\mathrm{rot}},
\end{equation}
where \(\Omega_{\mathrm{orb}}\) is the orbital frequency. We note that, in this article, we adopt the sign convention for \(k\) used by \citet{CAFeiN2013}, whereas \citet{gyre_tides2023} adopt the opposite convention. We also adopt the convention that \(\sigma_{m,k}\) is positive. If a negative value is obtained, the equivalent tidal component can be described by changing the signs of both \(m\) and \(k\). With our sign conventions, \(km<0\) corresponds to prograde waves in the inertial frame, whereas \(km>0\) corresponds to retrograde waves in the inertial frame. Prograde waves in the inertial frame generally have larger amplitudes because their Hansen coefficients are larger. In the co-rotating frame, \(m>0\) corresponds to retrograde waves, while \(m<0\) corresponds to prograde waves. Throughout this paper, unless stated otherwise, the terms ``prograde'' and ``retrograde'' refer to the co-rotating frame, which is the relevant frame for the mode physics.
\\~\\In the case of resonance locking, however, the long-term evolution is controlled by the forcing frequency that maintains the lock. We therefore restrict the analysis in this section to this single resonant component and use the asymptotic approximation \citep{Tassoul1980} to derive its impact on the orbital and rotational parameters of the binary system. Including the effect of rotation through the traditional approximation \citep{Lee1997,Townsend2005,Bouabid2013}, the asymptotic expression for the corotating frequency of an oscillation mode is
\begin{equation}\label{eq_asymptotic_approx}
\sigma=\dfrac{\sqrt{\lambda}\int N/r\ \mathrm{d}r}{\pi(n+1/2)}=\dfrac{\sqrt{\lambda}\mathcal{N}}{\pi(n+1/2)},
\end{equation}
where $\lambda$ is the eigenvalue associated with the Hough function of degree $\ell,m$ for a given spin parameter $\nu_{\mathrm{H}}$ ($\nu_{\mathrm{H}}=2\Omega_{\mathrm{rot}}/\sigma$), $N$ is the Brunt-Väisälä frequency, $r$ is the stellar radial coordinate, and $n$ is the radial order of the mode. In general, Hough functions and their eigenvalues $\lambda$ are obtained by solving Laplace's tidal equation as an eigenvalue problem, see for example \citep{Unno1989,Lee1997, Townsend2003}.  We define $\mathcal{N}$ as
\begin{equation}
\mathcal{N}=\int \dfrac{N}{r} \mathrm{d}r.
\end{equation}
During resonance locking, the rate of change of the forcing frequency, obtained by differentiating \autoref{eq_forcing_frequency}, is equal to that of the corotating oscillation frequency. The locking condition is therefore
\begin{equation}\label{eq_full_locking_condition}
\dfrac{\mathrm{d}\sigma}{\mathrm{d}t}=\dfrac{\partial \sigma}{\partial t}+\dfrac{\partial \sigma}{\partial \Omega_{\mathrm{rot}}}\dfrac{\mathrm{d}\Omega_{\mathrm{rot}}}{\mathrm{d} t}=k\dfrac{\mathrm{d}\Omega_{\mathrm{orb}}}{\mathrm{d} t}+m\dfrac{\mathrm{d}\Omega_{\mathrm{rot}}}{\mathrm{d} t}.
\end{equation}
We decompose the time derivative of the stellar rotation rate into tidal and evolutionary components, as in \autoref{eq_decomposition_rotation}. Using Kepler's third law, we transform $\mathrm{d}a/\mathrm{d}t$ in \autoref{eq_secular_a} into $\mathrm{d}\Omega_{\mathrm{orb}}/\mathrm{d}t$. Taking the ratio of Eqs.~\ref{eq_secular_a} and~\ref{eq_secular_rot} then eliminates their common dependence on $\mathrm{Im}(F_{\ell,m,k})$, which is highly sensitive to the resonance detuning. Substituting the resulting relation into \autoref{eq_full_locking_condition} and isolating the tidal contribution to the stellar spin evolution gives
\begin{equation}\label{eq_min_domegaR}
\left(\dfrac{\mathrm{d}\Omega_{\mathrm{rot}}}{\mathrm{d} t}\right)_{\mathrm{tides}}=\dfrac{-(m  - \partial \sigma/\partial \Omega_{\mathrm{rot}} )\left(\dfrac{\mathrm{d}\Omega_{\mathrm{rot}}}{\mathrm{d} t}\right)_{\mathrm{evo}}+\partial \sigma/\partial t}{m+ k  \mathrm{d}\Omega_{\mathrm{orb}}/\mathrm{d} \Omega_{\mathrm{rot}} - \partial \sigma/\partial \Omega_{\mathrm{rot}} }.
\end{equation}
\\~\\Similarly, the semi-major-axis and eccentricity evolution during resonance locking to a single oscillation mode are given by
\begin{equation}\label{eq_locking_a}
\dfrac{\mathrm{d}a}{\mathrm{d} t}= \dfrac{2}{3}\left(\dfrac{a^5}{G(M_1+M_2)}\right)^{1/2}  \dfrac{(m  - \partial \sigma/\partial \Omega_{\mathrm{rot}} )\left(\dfrac{\mathrm{d}\Omega_{\mathrm{rot}}}{\mathrm{d} t}\right)_{\mathrm{evo}}+\partial \sigma/\partial t}{ k  + (m- \partial \sigma/\partial \Omega_{\mathrm{rot}}) (\mathrm{d} \Omega_{\mathrm{orb}}/\mathrm{d}\Omega_{\mathrm{rot}})^{-1}}.
\end{equation}
\begin{equation}\label{eq_locking_e}
\dfrac{\mathrm{d} e}{\mathrm{d} t}=\dfrac{(m  - \partial \sigma/\partial \Omega_{\mathrm{rot}} )\left(\dfrac{\mathrm{d}\Omega_{\mathrm{rot}}}{\mathrm{d} t}\right)_{\mathrm{evo}}+\partial \sigma/\partial t}{ k (\mathrm{d}\Omega_{\mathrm{orb}}/\mathrm{d}e) + (m- \partial \sigma/\partial \Omega_{\mathrm{rot}}) (\mathrm{d} \Omega_{\mathrm{rot}}/\mathrm{d}e)}.
\end{equation}

The expressions for $\partial \sigma/\partial t$ and $\partial \sigma/\partial \Omega_{\mathrm{rot}}$ are obtained by differentiating \autoref{eq_asymptotic_approx} with respect to time and rotation frequency, respectively:
\begin{equation}
\dfrac{\partial \sigma}{\partial t}= \frac{\partial \ln \mathcal{N}/\partial t}{1+x/2}\sigma,
\end{equation}
\begin{equation}
 \dfrac{\partial \sigma}{\partial \Omega_{\mathrm{rot}}}=\dfrac{1}{\nu_{\mathrm{H}}}\dfrac{x}{1+x/2},
\end{equation}
where
\begin{equation}
x=\dfrac{\mathrm{d} \ln \lambda}{\mathrm{d}\ln \nu_{\mathrm{H}}}.
\end{equation}
Finally, $(\mathrm{d}\Omega_{\mathrm{orb}}/\mathrm{d} \Omega_{\mathrm{rot}})$ is obtained by taking the ratio of Eqs.~\ref{eq_secular_a} and~\ref{eq_secular_rot}:
\begin{equation}
\dfrac{ \mathrm{d} \Omega_{\mathrm{orb}}}{\mathrm{d} \Omega_{\mathrm{rot, tides}}}=\dfrac{G^{(2)}_{\ell,m,k}(e)}{G^{(4)}_{\ell,m,k}(e)} \dfrac{M_1+M_2}{M_1M_2}I a^{-2},
\end{equation}
and, similarly,
\begin{equation}
\dfrac{\mathrm{d} \Omega_{\mathrm{orb}}}{\mathrm{d} e}=\frac{3}{2}\dfrac{G^{(2)}_{\ell,m,k}(e)}{G^{(3)}_{\ell,m,k}(e)}\left(\frac{G(M_1+M_2)}{a^3}\right)^{1/2},
\end{equation}
\begin{equation}
\dfrac{\mathrm{d} \Omega_{\mathrm{rot}}}{\mathrm{d} e}=\dfrac{G^{(4)}_{\ell,m,k}(e)}{G^{(3)}_{\ell,m,k}(e)}\left(\frac{Ga}{M_1+M_2}\right)^{1/2} \frac{M_1 M_2}{I},
\end{equation}
\autoref{eq_min_domegaR} provides a lower limit on the amplitude of dynamical tides required at a given stage of stellar evolution to establish stable resonance locking. If the peak amplitude of an unstable resonant mode exceeds this threshold, determined by the radial order of the mode and $k$, resonance locking can occur at the corresponding orbital parameters. In \autoref{eq_min_domegaR}, $m - \partial \sigma/\partial \Omega_{\mathrm{rot}} \gg k\,\partial \Omega_{\mathrm{orb}}/\partial \Omega_{\mathrm{rot}}$ because of the large difference between the stellar rotational and orbital moments of inertia. This allows us to neglect the dependence on $k$. Consequently, the fundamental properties of resonance locking do not depend strongly on the orbital parameters of the system, provided that the amplitude of the dynamical tide is sufficiently large, but rather on the oscillation and rotational characteristics of the star.
At this point, a distinction has to be made between prograde and retrograde modes.

\section{Locking with retrograde modes (relatively wide systems with rapidly rotating components)}\label{sect_locking_retrograde}
\subsection{Reference model for this study}
For this study, we adopted the same reference evolutionary sequence for a \(\gamma\)-Doradus star as in \citet{Fellay2026}. The sequence was computed with the Code Liégeois d'Évolution Stellaire \citep[CLÉS;][]{Scuflaire2008a} for a star with a mass of $1.5\,M_\odot$, near the lower end of the mass range of \(\gamma\)-Doradus stars. We adopted an initial hydrogen mass fraction of $X=0.72$, an initial solar metallicity of $Z=0.015$, a mixing-length parameter of $\alpha=2.0$, and a step-overshooting parameter of $\alpha_{\mathrm{over}}=0.2$. The evolution was followed until the central hydrogen mass fraction reached $X_C=0.05$, corresponding to an age of $2.7\,\mathrm{Gyr}$. The reference model parameters were chosen to ensure the presence of unstable modes at the zero-age main sequence and throughout a substantial part of the main-sequence evolution. In all stellar models, we used the AGSS09 abundances \citep{Asplund2009}, the FreeEOS equation of state \citep{Irwin2012}, the OPAL opacities \citep{Iglesias1996}, the atmospheric $T(\tau)$ relation from Model C of \citet{Atmosphere1981}, the mixing-length theory of convection as implemented by \citet{Cox1968}, and the nuclear reaction rates of \citet{Reaction2011}. The evolutionary tracks and pulsation properties of the models are publicly available in the model grid presented by \citet{Fellay2026b}. The model used in \autoref{subsect_upper_limit_retrograde} was selected from this evolutionary sequence at an age of $738\,\mathrm{Myr}$ as it lies near the centre of the instability strip.

\subsection{Upper orbital separation limit of validity of the prescription}\label{subsect_upper_limit_retrograde}
\begin{figure}[h]
\centering
\includegraphics[width=\hsize]{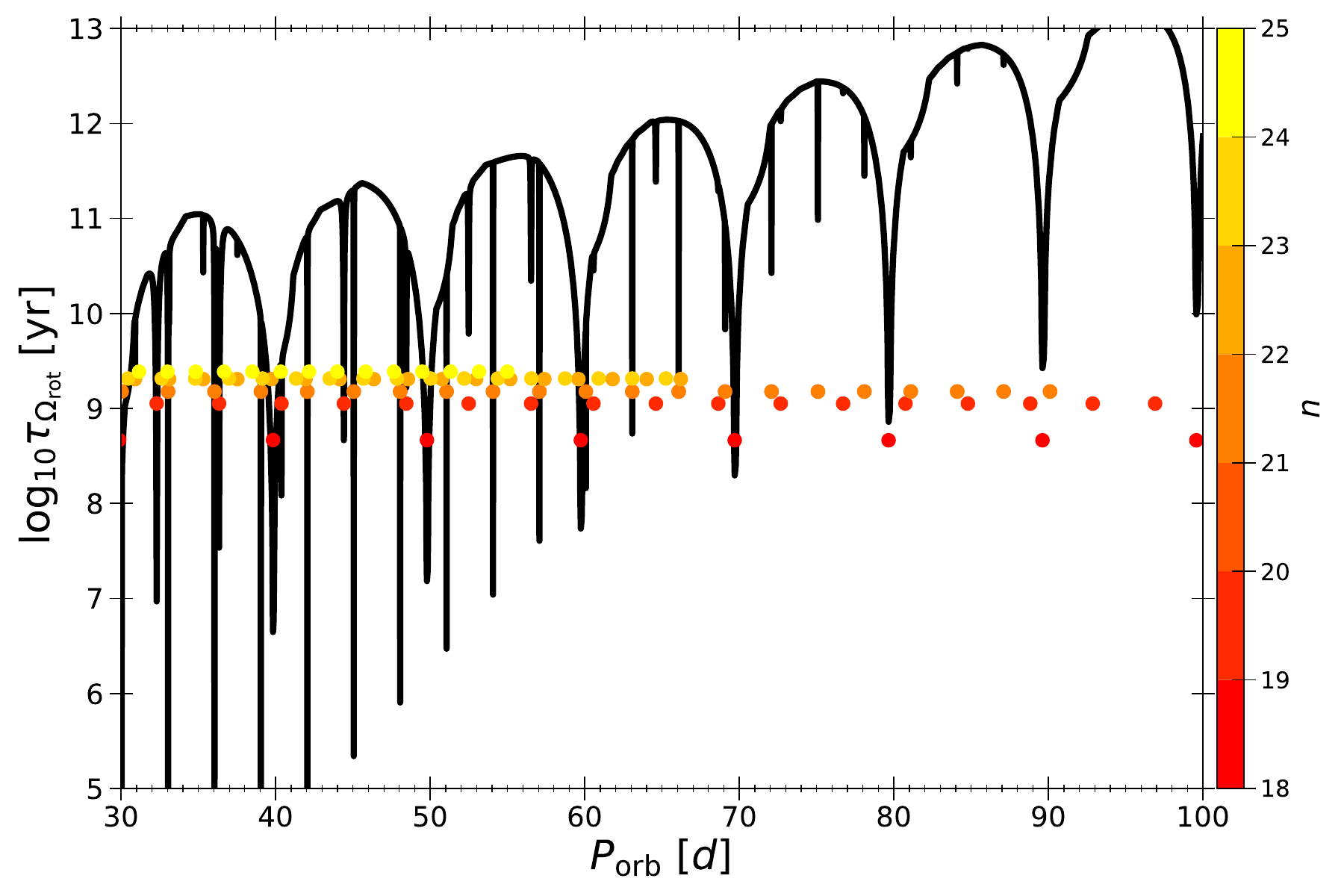}
\caption{Resonance-locking thresholds and tidal oscillation amplitudes in a twin $\gamma$-Doradus binary. The system consists of two $1.5\,M_\odot$ stars with an orbital eccentricity of $0.4$, a stellar rotation rate of $0.3\,\Omega_{\mathrm{crit}}$, and an age of $738\,\mathrm{Myr}$. The black line shows the amplitudes of tidally excited oscillations, while the coloured dots mark the theoretical thresholds for resonance locking. The colour of each dot indicates the radial order of the corresponding mode, and its vertical position gives the threshold amplitude.}
\label{fig.comp_amplitude_1M5_e0.4_wr0.3}
\end{figure}

\begin{figure}[h]
\centering
\includegraphics[width=\hsize]{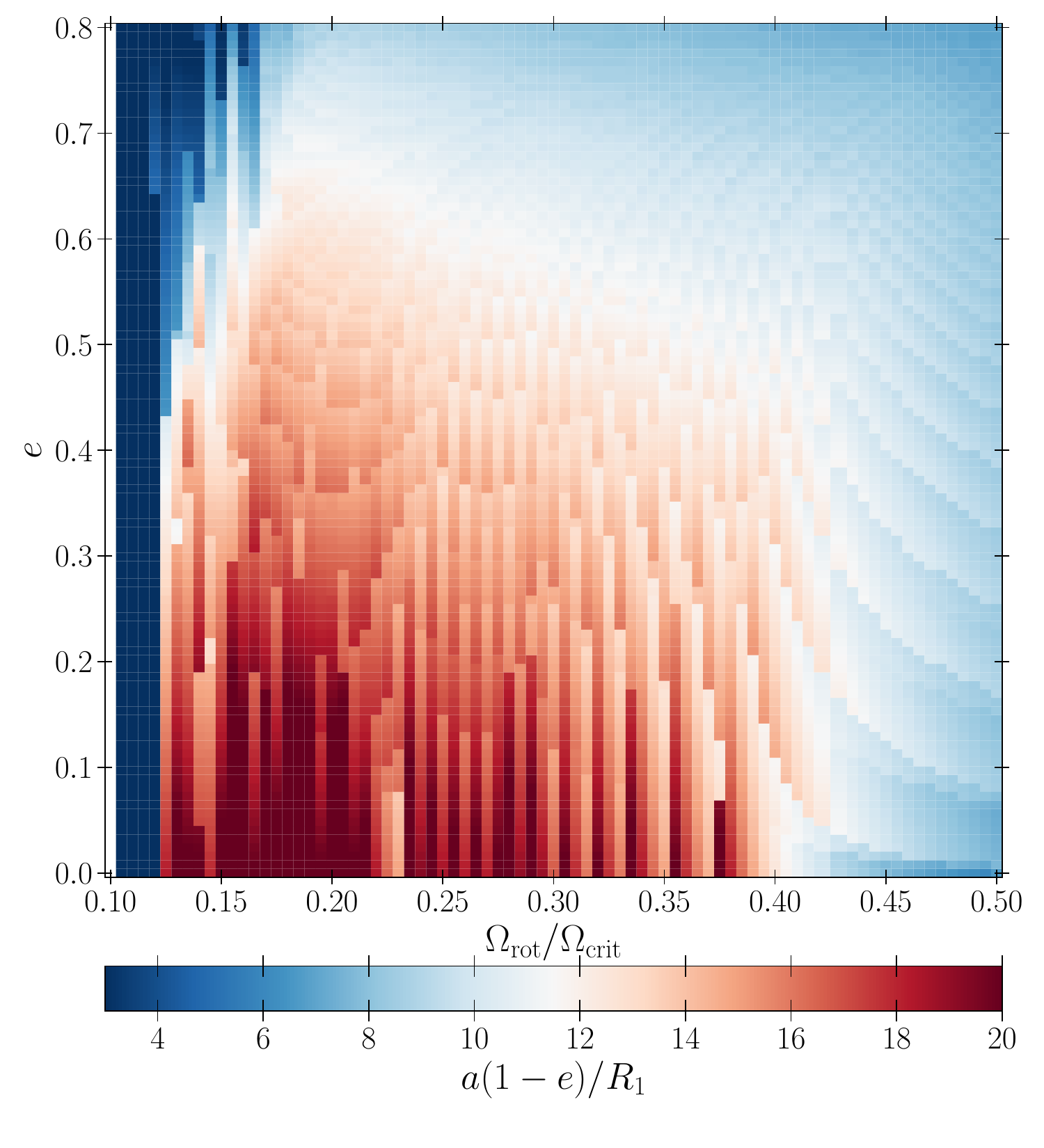}
\caption{Map of the conditions under which resonance locking can occur. The colour scale indicates the maximum periastron separation divided by the radius of the primary at which locking is possible, as a function of stellar rotation rate (x-axis) and eccentricity (y-axis), for the same $1.5\,M_\odot$ $738\,\mathrm{Myr}$ $\gamma$-Doradus star. For rotation rates below $12\%$ of the critical rotation, resonance locking with retrograde modes is not expected.}
\label{fig.map_amplitude_1M5}
\end{figure}

\begin{figure}[h]
\centering
\includegraphics[width=\hsize]{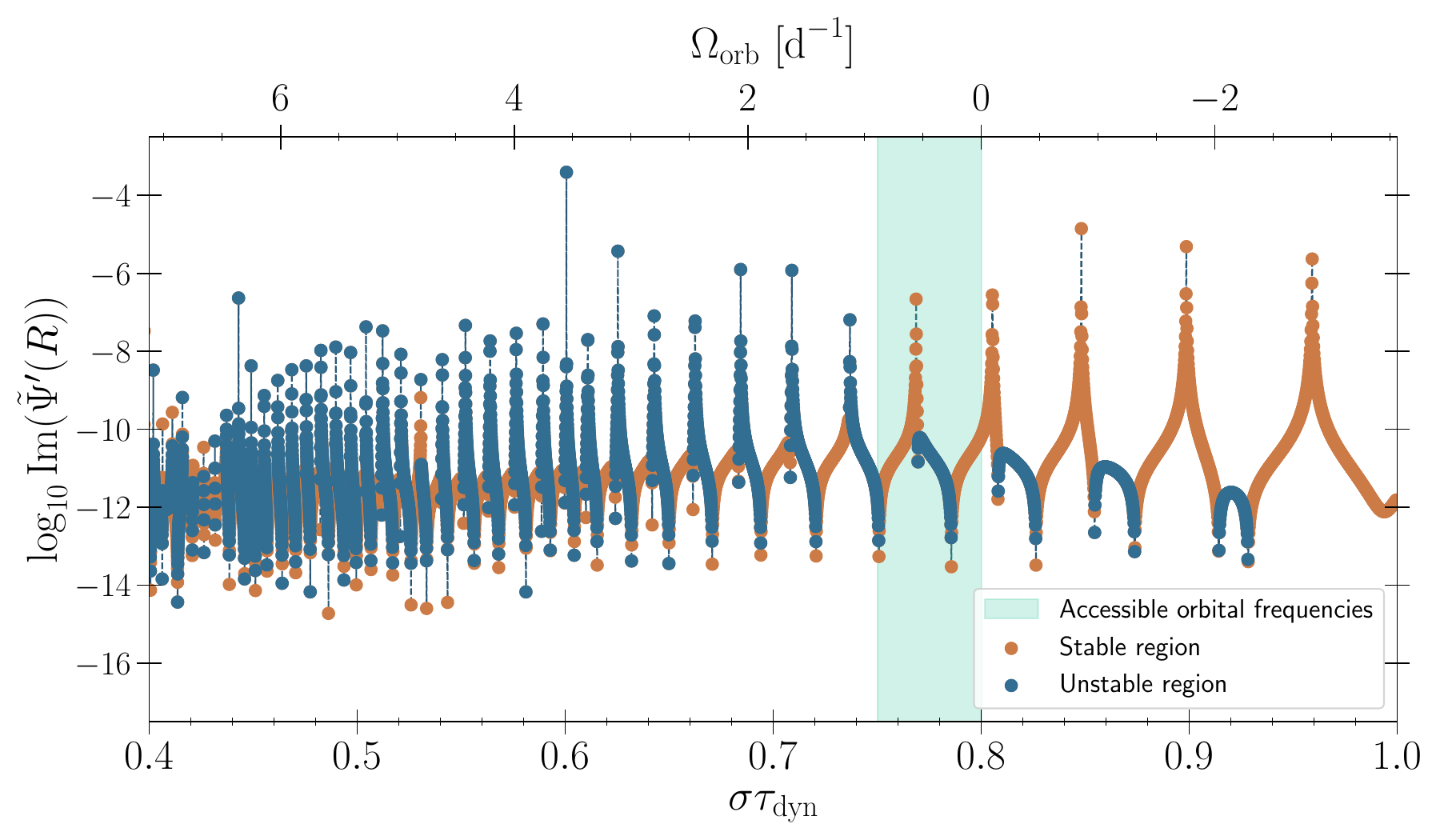}
\caption{Range of retrograde modes accessible at a rotation rate of $0.40\,\Omega_{\mathrm{crit}}$. The accessible range is restricted to orbital frequencies low enough to avoid contact and excessive tidal deformation. We adopt an arbitrary upper limit of $1\,\mathrm{d}^{-1}$ for the orbital frequency, although this value is likely already too high. The system is a twin $1.5\,M_\odot$ $\gamma$-Doradus binary at an age of $738\,\mathrm{Myr}$. For simplicity, we assume a circular orbit and adopt $k=-m=-2$. The y-axis shows the stellar response, quantified by the imaginary part of the dimensionless potential perturbation, as a function of the dimensionless forcing frequency on the lower x-axis and the orbital frequency on the upper x-axis. Blue dots represent unstable modes, which are the only retrograde modes capable of inducing resonance locking, as discussed above and by \citet{Fellay2026}. Orange dots represent stable modes, which cannot induce this type of locking. Negative values on the orbital-frequency axis correspond to modes that are prograde in the orbital frame. These modes are not considered here because their orbital-coupling coefficients, $G^{(2)}_{\ell,m,k}(e)$, $G^{(3)}_{\ell,m,k}(e)$, and $G^{(4)}_{\ell,m,k}(e)$, are very small, resulting in low tidal amplitudes.}
\label{fig.range_accessible_modes_0.4}
\end{figure}

\begin{figure}[h]
\centering
\includegraphics[width=\hsize]{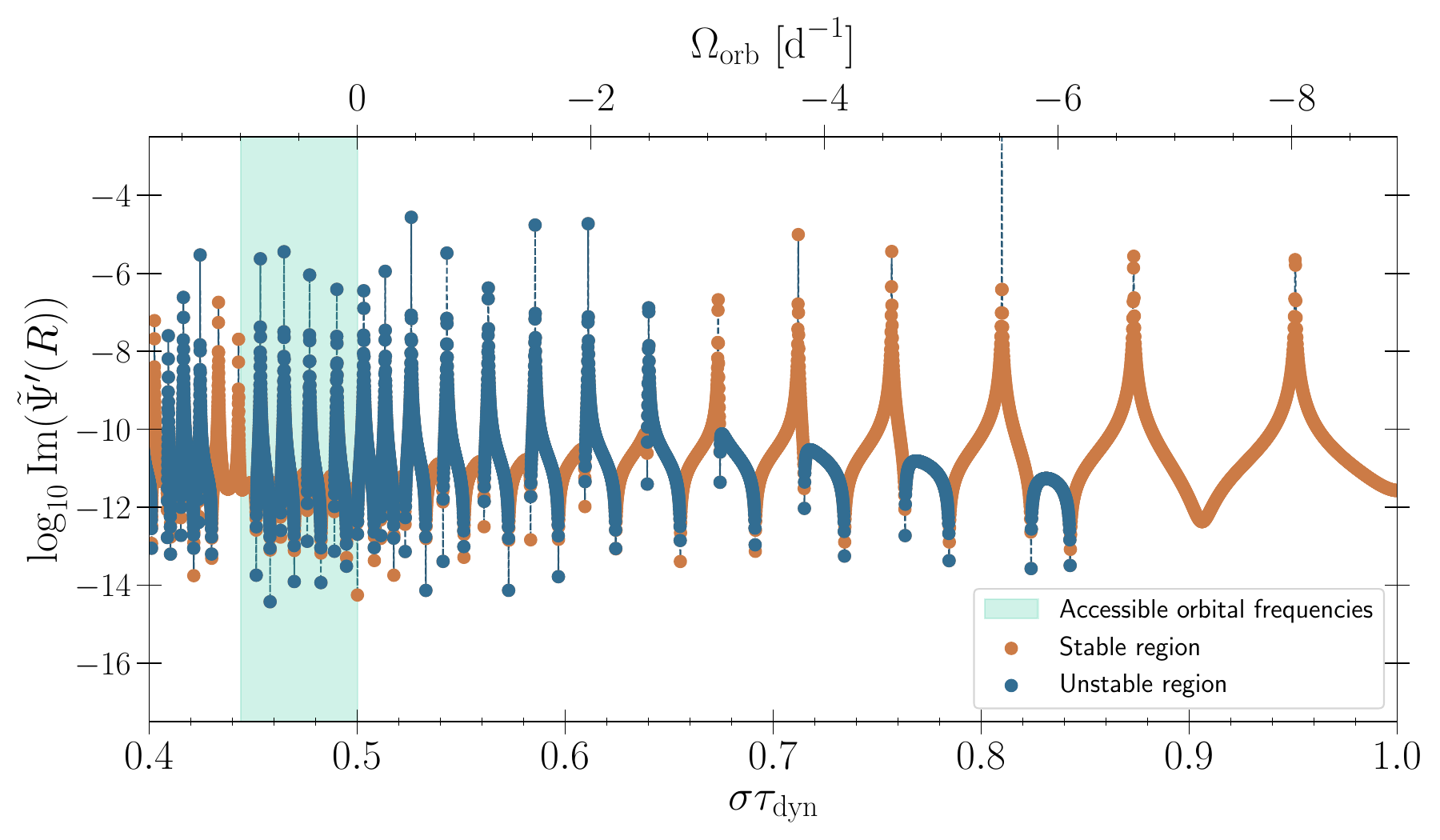}
\caption{Same as \autoref{fig.range_accessible_modes_0.4}, but for a lower rotation rate of $0.25\,\Omega_{\mathrm{crit}}$. At this rotation rate, the accessible modes are more numerous and are predominantly unstable, making resonance locking possible over a wider range of forcing frequencies.}
\label{fig.range_accessible_modes_0.25}
\end{figure}

In \autoref{subsect_rot_criterion_resonance_locking}, we established a criterion based on $\left(\mathrm{d}\Omega_{\mathrm{rot}}/\mathrm{d}t\right)_{\mathrm{tides}}$ to determine the amplitude of dynamical tides required for resonance locking to occur. In this section, we compare this criterion with the amplitudes of retrograde dynamical tides computed for different orbital and stellar parameters in order to determine the limiting orbital period up to which resonance locking is possible. For this comparison, we introduce the rotational-synchronisation timescale $\tau_{\Omega_{\mathrm{rot}}}$, defined as
\begin{equation}
\tau_{\Omega_{\mathrm{rot}}}=\Omega_{\mathrm{rot}}\left(\dfrac{\mathrm{d}\Omega_{\mathrm{rot}}}{\mathrm{d} t}\right)^{-1},
\end{equation}
which can be physically interpreted as the time taken for the system to reach synchronisation if the current tidal dissipation remains constant.
\\~\\For our $1.5\,M_\odot$ $\gamma$-Doradus model, applying this criterion is relatively straightforward, since a large fraction of the low-radial-order modes are unstable and are therefore expected to favour resonance locking. In \autoref{fig.comp_amplitude_1M5_e0.4_wr0.3}, we compare the amplitude required by the theoretical locking criterion with the amplitudes of the tidally excited oscillations obtained from scans of the oscillation spectrum. In these scans, we compute the orbital independant response of the star with our non adiabatic oscillation code MAD as a function of the forcing frequency before reconstructing the rate of change of the stellar rotation using the orbital properties of the binary system we assume (see \citet{Fellay2026} for the mathematical formulation).
For this comparison, we consider a system with an orbital eccentricity of $0.4$, an age of $738$ Myr, and a stellar rotation rate of $0.3\,\Omega_{\mathrm{crit}}$, $\Omega_{\mathrm{crit}}$ begin the stellar critical rotation defined as
\begin{equation}
\Omega_{\mathrm{crit}}=\sqrt{\dfrac{G M_1}{R_1^3}}.
\end{equation}
The results shown in \autoref{fig.comp_amplitude_1M5_e0.4_wr0.3} include the contribution of
$k\,\partial\Omega_{\mathrm{orb}}/\partial\Omega_{\mathrm{rot}}$
to \autoref{eq_min_domegaR}. As expected, however, the theoretical locking timescale criterion, indicated by the vertical positions of the dots, is independent of both $k$ and the orbital period. This is reflected in the constant required amplitude across orbital periods for a given radial order. By comparing the locking threshold with the actual amplitudes of the dynamical tides, we find that, for the particular system shown in \autoref{fig.comp_amplitude_1M5_e0.4_wr0.3}, resonance locking remains possible up to an orbital period of about $70$ days.  
\\~\\Using the same scan-based method, we repeated this comparison for stellar rotation rates ranging from $10\%$ to $50\%$ of the critical rotation rate, in steps of $0.5\%$, and for orbital eccentricities ranging from $0$ to $0.8$, in steps of $0.008$. For each scan, following the procedure illustrated in \autoref{fig.comp_amplitude_1M5_e0.4_wr0.3}, we determined the maximum orbital period for which resonance locking with retrograde modes remains possible. We then used the orbital properties of the system to calculate the corresponding maximum periastron separation. We report the maximum periastron separation because the maximum orbital period depends strongly on eccentricity, with longer periods being permitted at higher eccentricities. For the $1.5\,M_\odot$ $\gamma$-Doradus model at an age of $738\,\mathrm{Myr}$, \autoref{fig.map_amplitude_1M5} shows this maximum periastron separation for a locking to take place as a function of eccentricity and stellar rotation rate.
For the $1.5\,M_\odot$ $\gamma$-Doradus model, resonance locking is not expected for rotation rates below about $12\%$ of the critical rotation, because unstable modes that are retrograde in the corotating frame and prograde in the inertial frame become too scarce. At higher rotation rates, resonance locking remains possible, but its domain of occurrence changes. For rotation rates above about $40\%$ of the critical rotation, locking is expected only in systems with close periastron passages, approximately
\[
a(1-e)/R_1 \lesssim 12,
\] because the forcing then resonates mainly with low-radial-order stable modes when $\Omega_\star/\Omega_{\mathrm{crit}} \gtrsim 0.4$. 
By contrast, in the intermediate regime
\[
0.13 \lesssim \Omega_\star/\Omega_{\mathrm{crit}} \lesssim 0.40,
\] resonance locking can occur at much larger periastron separations, up to several tens of $R_1$. Although locking is therefore possible even in relatively wide systems, smaller separations are more favourable because they allow access to a larger number of high-amplitude modes capable of maintaining the lock. In \autoref{fig.map_amplitude_1M5}, the maximum separation at which resonance locking occurs is clearly non-monotonic and displays a stripe-like pattern. This behaviour arises because high-amplitude modes become accessible at low orbital frequencies when the stellar rotation frequency lies close to a free oscillation eigenfrequency. To illustrate this behaviour, \autoref{fig.range_accessible_modes_0.4} shows the range of modes that can be reached for a rotation rate of $0.40\,\Omega_{\mathrm{crit}}$, while \autoref{fig.range_accessible_modes_0.25} shows the corresponding range for a rotation rate of $0.25\,\Omega_{\mathrm{crit}}$.
The stripe-like pattern seen in \autoref{fig.map_amplitude_1M5} is directly related to the accessibility of these modes. Each time an unstable mode becomes reachable at low orbital frequency, resonance locking can in principle occur in that low-frequency regime. As the rotation rate increases, the accessible modes become sparser. Since high-order g-modes are approximately equally spaced in period \citep{Tassoul1980}, new unstable modes are encountered less frequently as the rotation increases, which explains the stripe-like pattern seen in the high-rotation regime of \autoref{fig.map_amplitude_1M5}. In addition, whereas at low rotation rates the accessible modes are mostly unstable, at higher rotation rates the accessible modes increasingly become stable. This explains why resonance locking can no longer be found at low orbital frequencies once the rotation rate approaches roughly $0.40\,\Omega_{\mathrm{crit}}$. Overall, these results suggest that resonance locking, and more generally inverse-tide effects involving unstable modes, may influence the evolution of $\gamma$-Doradus binaries over a broad range of orbital separations, including systems that are not extremely compact.

\subsection{Lower orbital separation limit of validity of the prescription}\label{subsect_lower_limit_retrograde}
As shown in \autoref{subsect_upper_limit_retrograde}, the upper orbital-frequency limit of the prescription, and therefore the region where resonance locking is expected to occur, depends on the amplitude of the dynamical tide in the high-frequency regime. For the low-frequency boundary, the relevant condition is instead whether the forcing frequency can reach the part of the mode spectrum containing unstable retrograde modes. This is therefore the criterion adopted here to estimate whether resonance locking with retrograde modes can occur. In the co-rotating frame, unstable \(\gamma\)-Doradus modes are typically found with periods between \(0.5\) and \(3\) days, as supported by observations and ensemble asteroseismic studies \citep{Li2019,Li2020,Garcia2022}. This corresponds to frequencies in the range
\[
\frac{1}{3}\,{\rm d}^{-1} \lesssim \sigma \lesssim 2\,{\rm d}^{-1}.
\]
We use this range as a simple criterion to estimate the domain of validity of the prescription. A more accurate treatment requires computing the instability range directly, as done in the following section. The aim of the present section is instead to identify the approximate range of orbital and rotational periods for which resonance locking with retrograde modes may occur and have an observable impact on binary evolution. Expressing the orbital frequency as a function of the forcing frequency and the rotation frequency gives
\begin{equation}\label{eq_omega_orb_forcing}
\Omega_{\mathrm{orb}}
=
\dfrac{1}{k}
\left(
\sigma - m\Omega_{\mathrm{rot}}
\right).
\end{equation}
For locking with retrograde modes, we consider \(m=2\). In the circular or nearly circular case, the dominant tidal component satisfies \(m=-k=2\), so that \(k=-2\). More generally, for \(k<0\)\footnote{Combined with the requirement of a positive orbital frequency, this condition ensures that the modes are prograde in the inertial frame and retrograde in the corotating frame, for which the Hansen coefficients are not negligible.}, the condition that the forcing frequency lies within the unstable-mode range gives (in d$^{-1}$):
\begin{equation}
\dfrac{2}{\vert k \vert}
\left(\Omega_{\mathrm{rot}}-\dfrac{1}{6}\right)
\geq
\Omega_{\mathrm{orb}}
\geq
\dfrac{2}{\vert k \vert}
\left(\Omega_{\mathrm{rot}}-1\right).
\end{equation}
Since the orbital frequency must be positive, this condition implies that at least part of the unstable retrograde-mode range can be accessed only if
\[
\Omega_{\mathrm{rot}}>\frac{1}{6}\,{\rm d}^{-1},
\]
or equivalently if the rotation period is shorter than approximately six days. In the circular case, \(m=-k=2\), the above condition reduces (in d$^{-1}$) to:
\begin{equation}
\Omega_{\mathrm{rot}}-\frac{1}{6}
\geq
\Omega_{\mathrm{orb}}
\geq
\Omega_{\mathrm{rot}}-1.
\end{equation}
This provides the approximate range of orbital frequencies, or equivalently orbital periods, in which resonance locking with retrograde modes can occur, before accounting for the amplitude constraint discussed in the previous subsection. For example, in the case of a twin \(1.5\,M_\odot\) \(\gamma\)-Doradus binary at an age of \(738\) Myr and rotating at \(20\%\) of the critical rotation rate, corresponding to a rotation frequency of \(1.13\,{\rm d}^{-1}\), this criterion gives an orbital-frequency range of approximately \(0.13--0.96\,{\rm d}^{-1}\). This corresponds to orbital periods between roughly \(1.0\) and \(7.7\) days. The more general implications of this limit are discussed in \autoref{sect_discussion}.

\subsection{Comparison with the full impact of resonance locking}
\begin{figure}[h]
\centering
\includegraphics[width=\hsize]{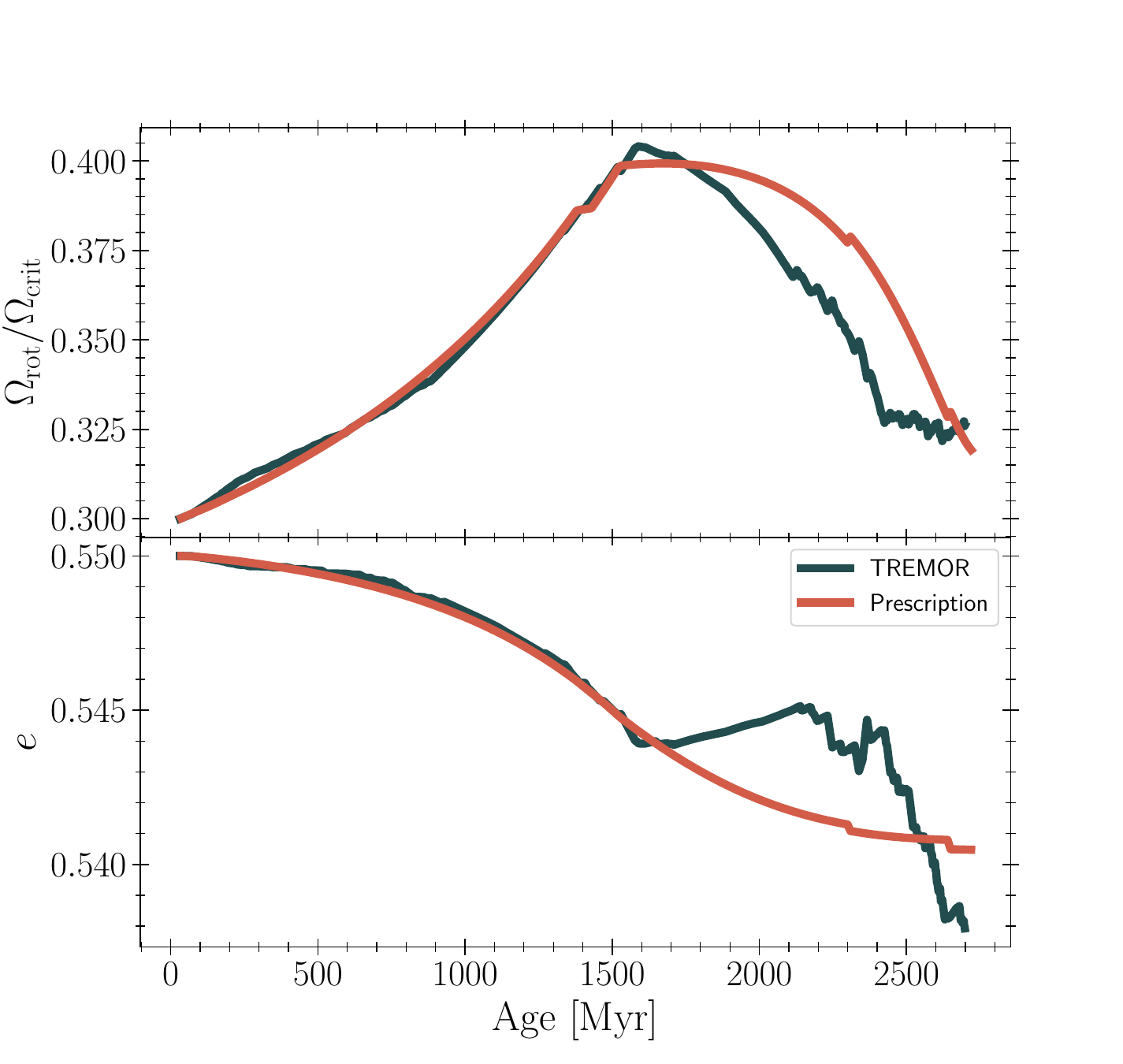}
\caption{Comparison between the stellar-rotation and eccentricity evolution obtained from the full \texttt{TREMOR} integration (blue curve) and the resonance-locking prescription (orange curve) for a twin binary composed of two $1.5\,M_\odot$ $\gamma$-Doradus stars. The system has the same orbital properties as in Fig.~5 of \citet{Fellay2026}, but with an initial stellar rotation rate of $0.3\,\Omega_{\mathrm{crit}}$.}
\label{fig.comp_1M5_e0.55_rot0.3}
\end{figure}

\begin{figure}[h]
\centering
\includegraphics[width=\hsize]{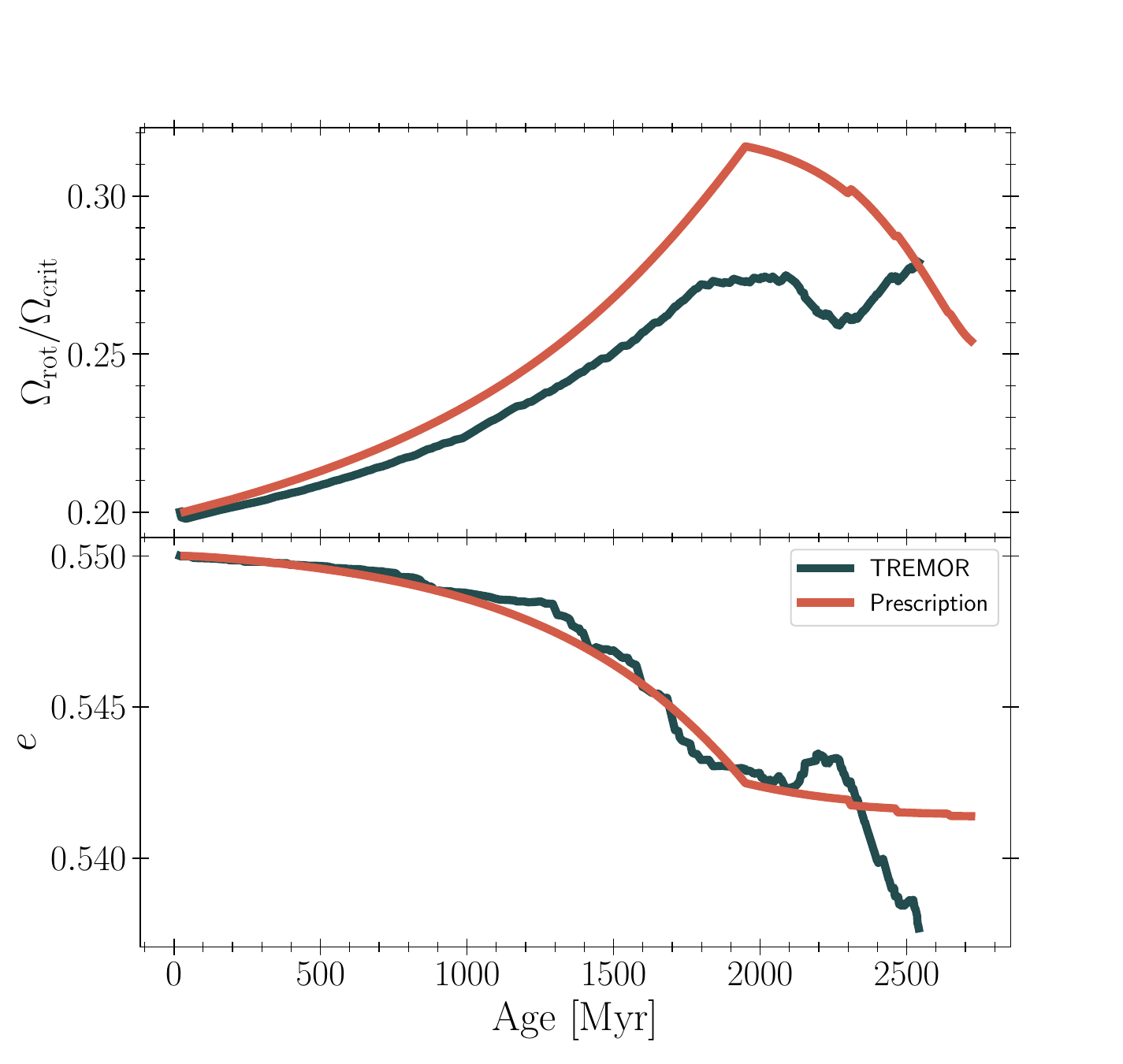}
\caption{Comparison between the stellar-rotation and eccentricity evolution obtained from the full \texttt{TREMOR} integration (blue curve) and the resonance-locking prescription (orange curve) for a twin binary composed of two $1.5\,M_\odot$ $\gamma$-Doradus stars. The system has the same orbital properties as in Fig.~5 of \citet{Fellay2026}, but with an initial stellar rotation rate of $0.2\,\Omega_{\mathrm{crit}}$.}
\label{fig.comp_1M5_e0.55_rot0.2}
\end{figure}

In \citet{Fellay2026}, we used \texttt{TREMOR}, an evolutionary integrator coupled to the non-adiabatic oscillation code \texttt{MAD} \citep{Dupret2001,Dupret2002,Dupret2005}, to follow the effects of dynamical tides associated with multiple simultaneous forcing frequencies throughout the main-sequence evolution of intermediate-mass stars. These full integrations showed that resonance locking can strongly modify the rotational and orbital evolution of $\gamma$-Doradus binaries and identified two main regimes: rapidly rotating systems, in which locking occurs mainly with retrograde modes, and close-in, slowly rotating systems, in which locking with prograde modes can occur (see \autoref{sect_prograde_modes}). In \autoref{subsect_rot_criterion_resonance_locking}, we derived a simplified prescription for the tidal evolution of a system during resonance locking. In this section, we apply this prescription to the retrograde-mode regime and compare the resulting evolution with the full \texttt{TREMOR} integrations. The prograde-mode regime is investigated separately in \autoref{sect_prograde_modes}. This comparison allows us to determine whether the prescription reproduces the main secular effects of resonance locking without requiring the full non-adiabatic tidal response to be computed at each evolutionary step. Here, we compare those full integrations with the prescription derived in the previous section in order to determine which aspects of the evolution can be captured by the simplified model and where a complete integration of the dynamical tide remains necessary.
\\~\\For the high-eccentricity systems presented by \citet{Fellay2026}, we integrate the prescription for the evolution of the stellar rotation, semi-major axis, and eccentricity. In the regime where resonance locking does not occur, we assume that the tidal dissipation is dominated by non-resonant dynamical tides described by the formalism of \citet{Zahn1975,Zahn1977}. We consider a twin binary system composed of two $1.50\,M_\odot$ $\gamma$-Doradus stars, with an initial orbital period of $12.28$~days, an initial eccentricity of $0.55$, and an initial rotation rate for each star equal to $30\%$ of the critical rotation, as in \citet{Fellay2026}. The comparison between the integration of the prescription and the results obtained with \texttt{TREMOR} is shown in \autoref{fig.comp_1M5_e0.55_rot0.3}.
By comparing the prescription with the results of the full \texttt{TREMOR} integration, we find that the two approaches agree reasonably well during the first part of the evolution, which corresponds to the resonance-locking phase. During this stage, the prescription is able to reproduce both the rotational and eccentricity evolution of the system. We also note that, by combining the results of the non-adiabatic computations with the predicted range of unstable modes, we are able to identify the part of the evolution during which resonance locking is expected to occur. When the locking phase ends, at around $1.5$~Gyr, differences begin to appear between the full integration and the prescription. In this later stage of the evolution, we adopt the prescriptions of \citet{Zahn1975,Zahn1977} to describe the tidal evolution. However, the full integration of tidally excited oscillations predicts a significantly stronger tidal interaction than the one obtained from the \citet{Zahn1975,Zahn1977} formalism. This discrepancy is likely related to the regime of validity of these prescriptions, which were derived for low forcing frequencies. In the systems considered here, the higher forcing frequencies produce a tidal response that lies outside this regime and is therefore more accurately captured by \texttt{TREMOR}. As a consequence, at the end of the evolution, the prescription underestimates the changes in both eccentricity and stellar rotation. These corrections can only be recovered through a full integration of the effect of tidally excited oscillations.
\\~\\In the case of more slowly rotating stars, we also explored the differences between the full integration and the prescription. We considered a twin binary system composed of two $1.50\,M_\odot$ $\gamma$-Doradus stars with the same initial orbital properties as in the previous case, but with an initial rotation rate of $20\%$ of the critical rotation. The comparison between the full integration and the prescription is shown in \autoref{fig.comp_1M5_e0.55_rot0.2}.
As illustrated in \autoref{fig.comp_1M5_e0.55_rot0.2}, the differences between the prescription and the full integration are expected to be larger over the whole evolution. In the case of slower rotation, resonance locking appears to be less efficient, or at least less frequent. This can be understood from the fact that such a rotation rate only allows the system to reach unstable modes with very high radial orders. As a result, a large fraction of the theoretical instability strip becomes inaccessible, making locking less likely. Consequently, when the system becomes unlocked, it typically takes longer to re-enter a locking phase, and dissipative tides are able to act during these intervals. As in the faster-rotation case, we also observe strong discrepancies at the end of the evolution, owing to the growing influence of dissipative tides and to the limitations of the prescription of \citet{Zahn1975,Zahn1977} in the high-forcing-frequency regime. This behaviour is expected, since these prescriptions were not derived for such conditions. Overall, the comparison confirms that, for slowly rotating systems, the simplified prescription is less able to reproduce the detailed secular evolution obtained from the full \texttt{TREMOR} integration.

\section{Locking with prograde modes (close binaries with slowly rotating components)}\label{sect_prograde_modes}
The key difference with respect to retrograde modes is that, in this case, resonance locking occurs near amplitude dips in the dynamical-tide spectrum. The main effect of this locking is to impose a slow secular evolution of the stellar rotation, thereby preventing the rapid synchronisation that would otherwise be expected in close-in systems \citep{Fellay2026}. This type of resonance locking occurs in close binaries, where the forcing frequency is dominated by the orbital frequency. In this regime, the stellar rotation is expected to remain slow throughout the evolution, as this slow rotation is maintained by the locking itself.
 
\subsection{Limit of validity of the prescription}\label{subsect_limit_prograde}
Because the tidal forcing is strong in these close-in systems, the absolute amplitudes of the tidally excited oscillations are not expected to be the limiting factor for resonance locking, even when the lock occurs near a local minimum of the response spectrum. Instead, as discussed in \autoref{subsect_lower_limit_retrograde}, the relevant limit is set by the range of forcing frequencies that can reach the region of unstable modes. As already mentioned, unstable modes in the co-rotating frame are found in the frequency range
\[
\frac{1}{3}\,{\rm d}^{-1} \lesssim \sigma \lesssim 2\,{\rm d}^{-1}.
\]
Expressing the orbital frequency as a function of the forcing frequency and the rotation frequency, as in \autoref{eq_omega_orb_forcing}, gives (in d$^{-1}$), for prograde modes,
\begin{equation}
\dfrac{2}{k}
\left(\Omega_{\mathrm{rot}}+\dfrac{1}{6}\right)
\leq
\Omega_{\mathrm{orb}}
\leq
\dfrac{2}{k}
\left(\Omega_{\mathrm{rot}}+1\right).
\end{equation}
For circular or nearly circular close binaries, the dominant tidal component is the quadrupolar term with \(m=-2\) and \(k=2\), i.e. \(m=-k\). The condition for resonance locking with unstable prograde modes therefore becomes (in d$^{-1}$):
\begin{equation}
\Omega_{\mathrm{rot}}+\frac{1}{6}
\leq
\Omega_{\mathrm{orb}}
\leq
\Omega_{\mathrm{rot}}+1.
\end{equation}
In the limiting case where the rotational frequency is negligible, this corresponds to orbital periods between approximately one and six days. This range therefore defines the approximate observational window in which resonance locking with prograde modes, or at least its rotational signature, may be detectable. It also indicates the regime in which the prescription is expected to be applicable, provided that the stars lie in the appropriate region of the HR diagram.

\subsection{Comparison with the full impact of resonance locking}
\begin{figure}[h]
\centering
\includegraphics[width=\hsize]{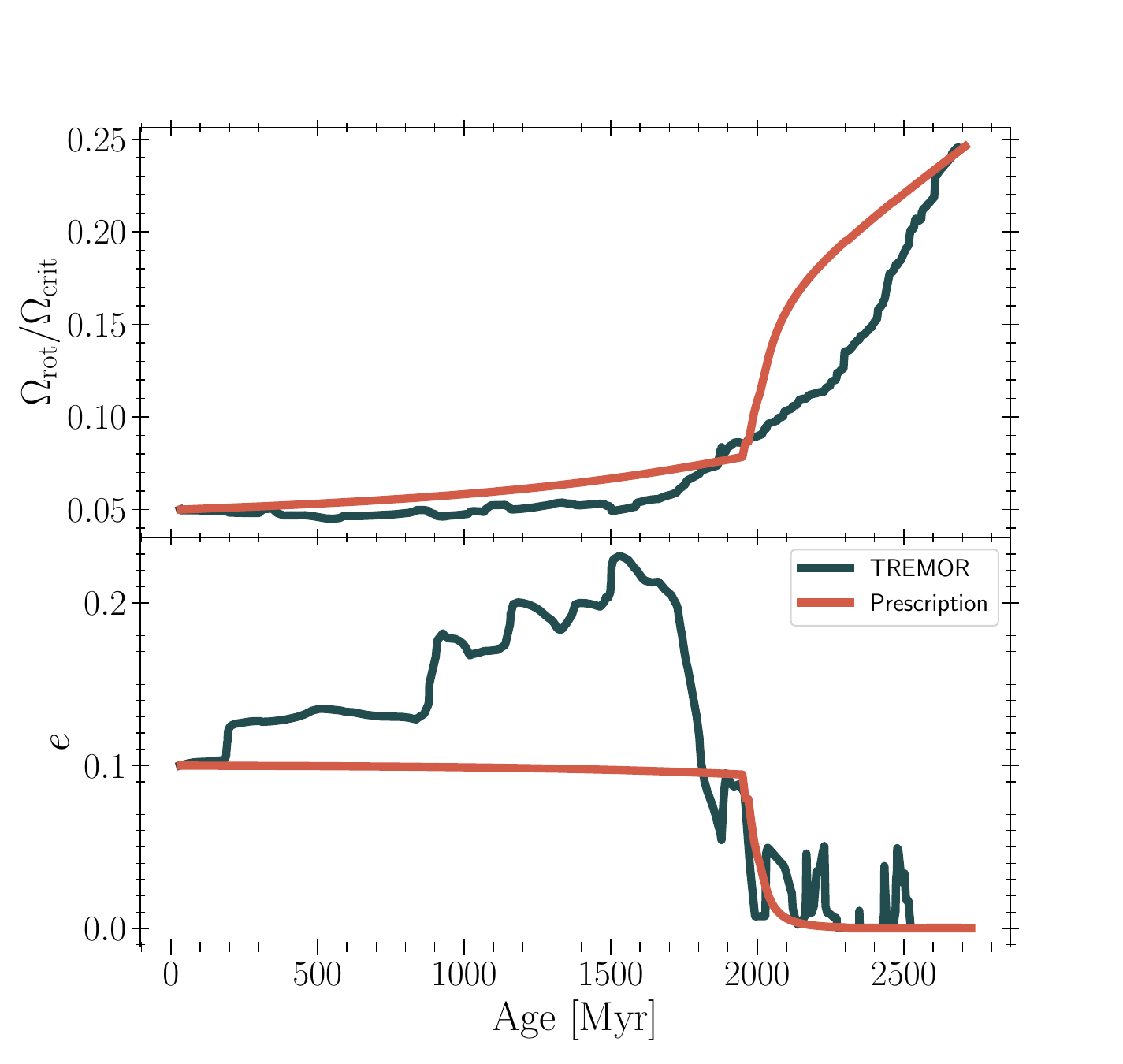}
\caption{Comparison between the stellar-rotation and eccentricity evolution obtained from the full \texttt{TREMOR} integration (blue curve) and the resonance-locking prescription (orange curve) for a twin binary composed of two $1.5\,M_\odot$ $\gamma$-Doradus stars. The system has an initial orbital eccentricity of $0.1$ and an initial stellar rotation rate of $0.05\,\Omega_{\mathrm{crit}}$.}
\label{fig.prescription_prograde}
\end{figure}

In the case of prograde modes, resonance locking is expected to occur in close binary systems with initially slow stellar rotation. When such locking takes place, the rotational evolution of the system is expected to be significantly slowed down with respect to that predicted by classical tidal prescriptions \citep{Zahn1975,Zahn1977}, as discussed in the previous section. Using the prescription, we therefore attempted to reproduce the results of the full time integration. The comparison between the full \texttt{TREMOR} calculations and the prescription is shown in \autoref{fig.prescription_prograde}.
As shown in \autoref{fig.prescription_prograde}, the prescription reproduces the rotational evolution reasonably well during the locking phase, which extends over approximately the first $1.5$~Gyr. In this phase, both the prescription and the full \texttt{TREMOR} integration predict a slow secular evolution of the stellar rotation, consistent with the idea that resonance locking can strongly reduce the efficiency of tidal spin evolution \citep{Fuller2017}. At later times, however, strong discrepancies appear between the \texttt{TREMOR} results and Zahn's prescription. This is not unexpected, since classical prescriptions are not designed to account for the detailed behaviour of resonance locking, and the strong tidal interactions in this regime amplify the differences between the two approaches. For the eccentricity evolution, the agreement is poorer over the whole integration. In that case, discrepancies build up progressively because of the repeated unlocking episodes experienced by the system. Each of these events has a significant impact on the eccentricity, owing to the strength of the tidal interaction in this regime. We therefore expect the prescription to perform less well when applied to the orbital evolution, especially in close systems where the tides are strong and where each unlocking episode can produce a non-negligible change in the orbital parameters. By contrast, for the stellar rotation, the prescription still captures the overall secular behaviour and reproduces the inhibition of the strong tidal spin evolution that would otherwise be expected in these systems.

\section{Discussion}\label{sect_discussion}
\begin{figure}[h]
\centering
\includegraphics[width=\hsize]{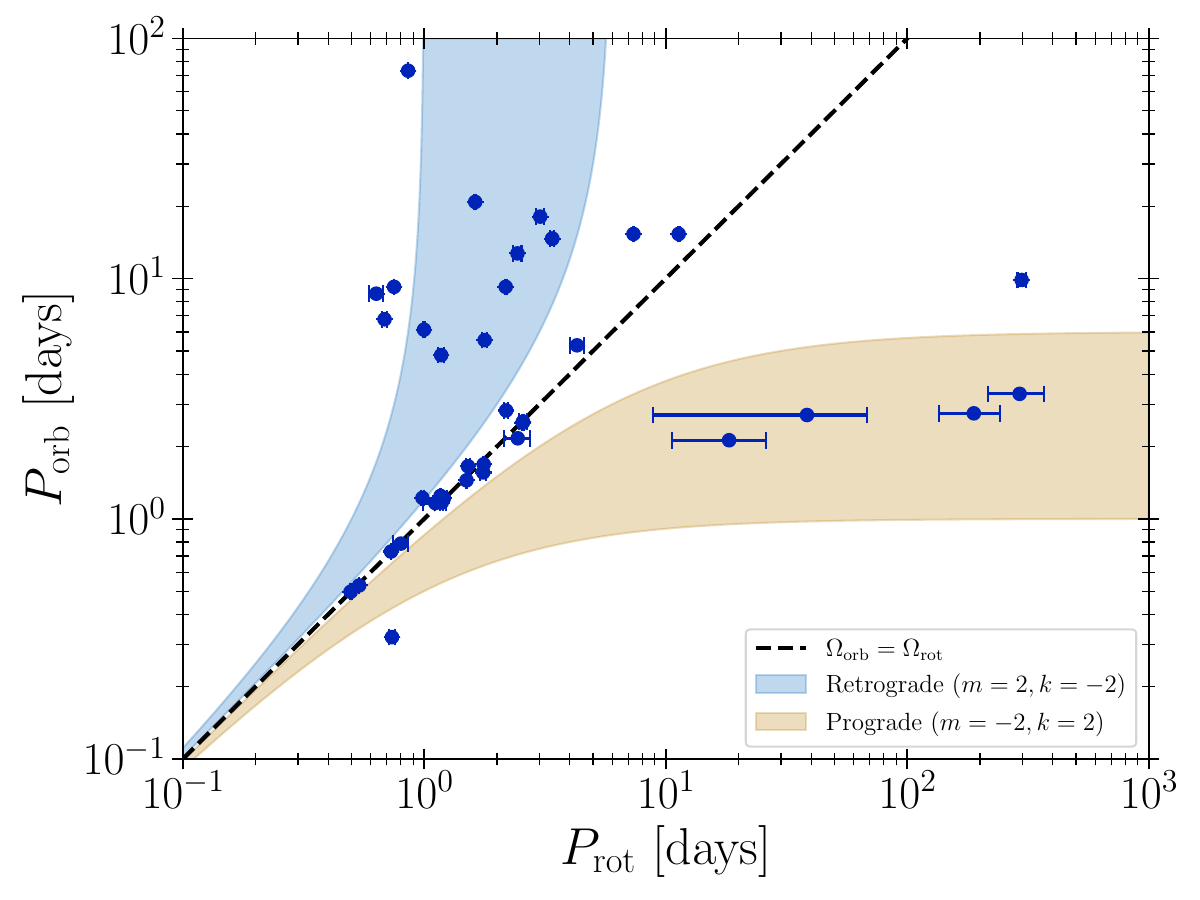}
\caption{
Observed orbital period versus near-core rotational period for close \(\gamma\)-Doradus binary systems. The observational sample is taken from the catalogue of \citet{Li2020}. The blue and yellow regions indicate the domains where resonance locking is expected to occur, and where strong spin--orbit asynchronism may therefore be maintained, according to the criteria derived in \autoref{subsect_lower_limit_retrograde} and \autoref{subsect_limit_prograde}. The amplitude criterion is not included in this figure, since it is model dependent.
}
\label{fig.range_locking}
\end{figure}

This article investigates the expected rotational and orbital evolution of binary systems undergoing resonance locking by introducing a prescription for the evolution of these quantities. Within this prescription, resonance locking is assumed to occur, and the regimes in which this assumption is expected to be valid are identified in \autoref{subsect_upper_limit_retrograde}, \autoref{subsect_lower_limit_retrograde}, and \autoref{subsect_limit_prograde}. Resonance locking remains a theoretically predicted phenomenon that is difficult to observe directly, although candidate systems have been identified \citep{FullerEtAl2017,Hambleton2018}. Its presence must therefore be inferred mainly from its long-term effects on binary evolution \citep{Witte1999b,Witte2001,Burkart2014,Fuller2017}. As shown by \citet{Fellay2026} and in the present work, resonance locking can maintain, and in some cases amplify, spin--orbit asynchronism. Its observational signature is thus expected to appear in asynchronous close binaries that are old enough for classical tidal dissipation to have synchronised them in the absence of locking.
\\~\\Observationally, however, only a limited number of \(\gamma\)-Doradus stars have been identified in close binary systems. The current sample is therefore still small and is mainly based on \textit{Kepler} observations. Starting from the sample of \(34\) \(\gamma\)-Doradus stars in binary systems identified by \citet{Li2020}, we can compare the predicted resonance-locking domain with the currently observed population of \(\gamma\)-Doradus stars in close binaries. This comparison is shown in \autoref{fig.range_locking}, where the observed systems are plotted together with the regions in which resonance locking is expected to occur, as derived in \autoref{subsect_lower_limit_retrograde} and \autoref{subsect_limit_prograde}.
As shown in the figure, about half of the \(36\) \(\gamma\)-Doradus stars in close binary systems are consistent with synchronous rotation, about one third rotate faster than synchronisation, and six systems are found to rotate more slowly. Many of the fast-rotating stars lie in the region where resonance locking is expected to occur. However, these systems could also be the result of initially rapidly rotating binaries that have not yet had time to synchronise, because tidal interactions are weaker at their relatively long orbital periods, typically above four days. Individual modelling of these systems is therefore required to determine their ages and to assess whether synchronisation should already have occurred. This would make it possible to establish whether their position in the diagram is genuinely anomalous and requires further investigation. The slowly rotating systems are also particularly interesting, since most of them lie in the region where resonance locking with prograde modes is expected to occur. In these systems, synchronisation would generally be expected to occur on relatively short timescales, of the order of a few hundred million years, as estimated by applying the synchronisation formalism of \citet{Zahn1975,Zahn1977} to \(\gamma\)-Doradus stars. Observing such slow rotation is therefore unexpected unless the systems are very young, although alternative mechanisms such as trapping in a Cassini state may also produce strongly subsynchronous rotation in hierarchical systems \citep{Felce2023}. Dedicated modelling of these stars is thus necessary to determine their ages and architectures and to assess whether they are plausible candidates for the long-term consequences of resonance locking.
\\~\\Improved constraints on these systems would be valuable for testing resonance locking and identifying its possible signatures. Even if resonance locking were eventually ruled out, a detailed characterisation of these systems would still provide important constraints on tidal dissipation in the regime of strong interactions. In particular, measuring both the near-core rotation, as already possible through asteroseismology of these stars \citep{Li2019,Li2020,Garcia2022,Saio2021}, and the surface rotation, for example from starspot modulation \citep{Lurie2017}, would help constrain where tidal dissipation acts inside the star. Such measurements could clarify how tidal synchronisation proceeds in strongly interacting systems, whether differential rotation or shear develops, and whether only the surface layers are affected \citep{Kallinger2017,Lurie2017}. In this context, future high-precision photometry from \textit{PLATO} will be particularly valuable for asteroseismology and measurements of surface rotation \citep{Rauer2025,Breton2024}. The Earth~2.0 (ET) mission is designed to revisit the original \textit{Kepler} field, providing a particularly long temporal baseline for many targets \citep{Ge2022}. For systems observed at multiple epochs, these data may make it possible to search for dynamical signatures such as orbital- or rotational-period variations, even if the likelihood of detecting such effects remains small.

\section{Conclusion}\label{sect_conclusion}
In this article, we derived an analytical prescription, based on the asymptotic approximation, to include the effect of inverse tides when resonance locking occurs. Under the assumption that a lock is established, this prescription makes it possible to predict the evolution of quantities such as \(a\), \(e\), and \(\Omega_{\mathrm{rot}}\). It should be emphasized that this prescription is not additive in the usual sense. When resonance locking takes place, the evolution is imposed by the locking condition itself, and additional tidal prescriptions, such as classical equilibrium tides or Zahn-type prescriptions, should not simply be added independently. If additional physical processes are to be included, the numerators of \autoref{eq_min_domegaR}, \autoref{eq_locking_a}, and \autoref{eq_locking_e} must be modified accordingly, in a way similar to what was done for the rotational evolution. Even if the net torque is modified by such additional processes, the rotational evolution is not expected to change significantly, since it is primarily imposed by the evolution of the locked mode. The quantity most likely to be affected is instead the amplitude of the dynamical tide required to maintain the locking. This amplitude should therefore be examined carefully before including any additional process, since resonance locking may no longer be possible if that process is too strong.
\\~\\Concerning the validity of the prescription, we showed that, for \(\gamma\)-Doradus stars, there are two regimes of orbital and rotational periods in which the forcing frequency is expected to reach the unstable-mode range. These regimes are described in \autoref{subsect_lower_limit_retrograde} and \autoref{subsect_limit_prograde}. For rapidly rotating stars, an additional condition must be imposed on the amplitude of the dynamical tide, or equivalently on the minimum periastron separation, as discussed in \autoref{subsect_upper_limit_retrograde}. In practice, when carrying out an integration, we recommend applying the locking prescription only if the considered forcing frequency lies within the unstable range of \(\gamma\)-Doradus modes and if the star is expected to be located inside the \(\gamma\)-Doradus instability strip in the HR diagram. A more robust approach is to use direct outputs from non-adiabatic oscillation calculations to verify these conditions. In particular, for all the prescriptions integrations carried out in this article we used the public grid of stellar \(\gamma\)-Doradus stellar models with non-adiabatic compuatations presented in \cite{Fellay2026b}.  At each timestep, these grids can be used to check whether the stellar model lies in the \(\gamma\)-Doradus instability strip and whether the forcing frequency falls within the corresponding unstable-mode range. This is the strategy adopted here to compare the prescription with the full integration results obtained with \texttt{TREMOR}. The agreement between the two approaches shows that this grid-based implementation provides an accurate and practical way to apply the prescription.
\\~\\When comparing the time evolution obtained from the prescription with the full dynamical-tide integration computed with \texttt{TREMOR}, we find that the orbital and rotational evolution are well reproduced as long as resonance locking is maintained. Thus, despite its simplicity and the fact that it only relies on quantities that can be computed without a non-adiabatic oscillation code, the prescription remains accurate within the locking regime. Outside this regime, however, simplified prescriptions become insufficient. As shown in this work, once resonance locking is lost, neither the present prescription nor Zahn's prescription reproduces the detailed orbital and rotational evolution obtained from the full calculation. In this case, a direct integration of the dynamical tide is required to capture the tidal evolution accurately.
\\~\\Finally, while this work focused on binaries composed of \(\gamma\)-Doradus stars, resonance locking should also be explored in the context of SPB stars. This is a natural extension, since both \(\gamma\)-Doradus and SPB stars are gravity-mode pulsators and can be studied within closely related seismic frameworks \citep{Townsend2005,Kallinger2017}. Although the prescriptions derived in this article remain formally applicable, their quantitative impact in the SPB regime still needs to be investigated in detail and compared with full \texttt{TREMOR} integrations. Preliminary numerical tests indicate that resonance locking can occur in SPB models. Nevertheless, a dedicated study is required to determine how this mechanism affects the orbital and rotational evolution of SPB binaries. In particular, the unstable-mode frequency ranges, the amplitude of the dynamical tide, and the relevant orbital and rotational regimes must be characterised specifically for SPB stars.

\section{Data availability}
The data required to implement or test these prescriptions, including tables of $\lambda$ as a function of the spin parameter, are available from the authors upon request. The grids of $\gamma$-Doradus stellar models, including the non-adiabatic oscillation calculations, are available at \url{https://doi.org/10.58119/ULG/F8OXLQ}.

\begin{acknowledgements} 
L.F. was supported by the Fonds de la Recherche Scientifique--FNRS (F.R.S.-FNRS) as a Research Fellow.
\end{acknowledgements}
\bibliography{Letter_prescription.bib}
\bibliographystyle{aa}
\appendix

\end{document}